\documentclass[%
 reprint,
 amsmath,amssymb,
 aps,
]{revtex4-2}

\usepackage{graphicx}
\usepackage{dcolumn}
\usepackage{bm}
\usepackage{subfigure}

\begin{document}

\preprint{}

\title{Ultimate bunch length and emittance performance of an MeV ultrafast electron diffraction apparatus with a DC gun and a multi-cavity SRF linac}

\author{A. Bartnik$^1$}
\author{C. Gulliford$^{3}$}
\author{G. H. Hoffstaetter$^{1,2}$}
\author{J. Maxson$^1$}

\affiliation{Cornell University$^1$, Ithaca/NY USA, Brookhaven National Laboratory$^2$, Upton/NY USA, Xelera$^3$, DE USA}





\date{\today}

\begin{abstract}
We present the design of a high repetition rate MeV energy ultrafast electron diffraction instrument based on a DC photoelectron gun and an SRF linac with multiple independently controlled accelerating and bunching cavities. The design is based on the existing Cornell photoinjector, which can readily be applied to the presented findings. Using particle tracking simulations in conjunction with multiobjective genetic algorithm optimization, we explore the smallest bunch lengths, emittance, and probe spot sizes achievable.  We present results for both stroboscopic conditions (with single electrons per pulse) and with $10^5$ electrons/bunch which may be suitable for single-shot diffraction images. In the stroboscopic case, the flexibility provided by the many-cavity bunching and acceleration allows for longitudinal phase space linearization without a higher harmonic field, providing sub-fs bunch lengths at the sample. Given low emittance photoemission conditions, these small bunch lengths can be maintained with probe transverse sizes at the single micron scale and below. In the case of $10^5$ electrons per pulse, we simulate state-of the art 5D brightness conditions: rms bunch lengths of 10 fs with 3 nm normalized emittances, while now permitting repetition rates as high as 1.3 GHz. Finally, to aid in the design of new SRF-based UED machines, we simulate the trade-off between the number of cavities used and achievable bunch length and emittance.  

\end{abstract}

\maketitle


\section{Introduction}
Electron diffraction \cite{Mourou:1982vvba, Sciaini:2011hiba} and microscopy \cite{Zewail:2006emba} with sub-picosecond temporal resolution have become invaluable tools for the discovery and characterization of a wide variety of phenomena that occur far from static equilibrium. These include phase transitions which have shown dramatic changes in material lattice symmetry \cite{Sie2019}, charge ordering \cite{Vogelgesang2017, Kogar2019}, electrical conductivity \cite{Morrison2014, Otto2018}, and may be a useful tool for continued study of light induced superconductivity \cite{Budden2021, Fausti2011, Mitrano2016}. 

Ultrafast electron diffraction (UED) has been performed with a variety of electron accelerator technology ranging from tabletop keV sources to high brightness photoinjectors with MeV energies. While more complex to build than their keV counterparts, MeV energies provide several advantages. The higher energy relativistically suppresses the space charge interaction \cite{reiser:book}, allowing for higher density bunches and/or more flexible electron optics setpoints. In term of diffraction performance, MeV electrons have deeper penetration into materials allowing for thicker samples, and furthermore, the flatter Ewald sphere of MeV electrons permits more efficient scattering into higher order Bragg peaks \cite{Zhu:2015ggba}.

MeV UED was pioneered with high brightness normal conducting RF photoguns \cite{Fu_2014, Li:2009kjba, Weathersby:2015jw, Gutierrez:2010cmba}, which can provide very high accelerating fields ($\sim 100$ MV/m) but at low repetition rates (hundreds of Hz or less). Recently, several MeV UED systems have been designed and constructed which are capable of operating at MHz repetition rates \cite{Filippetto:2016ueba, Feng_2015}, which is very useful for gas phase samples or for investigating solid state samples with non-destructive pump excitation energy density \cite{Lahme2014}. While usually operating with lower source gradients ($<50$ MV/m), these systems still retain the capability to generate sufficient charge for single shot diffraction pattern acquisition and furthermore have the ability to run in stroboscopic mode with charges as low as one electron per pulse \cite{Ji2019}.

In this work we present the design of an MeV-UED system based on a DC photoelectron gun and superconducting RF linac booster. This system is based on the Cornell SRF injector installed at the CBETA accelerator, originally designed for generation of 100 pC-scale bunches for high repetition rate light source applications \cite{ref:lowemitter}. It is capable of any repetition rate which is an integer divisor of 1.3 GHz.  While the source gradient is less than for RF guns (here up to $10$ MV/m), the injector is equipped with an atypically large number of transverse focusing elements (3 solenoids) and independently controlled 1.3 GHz RF cavities (5 SRF cavities and 1 normal conducting cavity). The larger number of elements is due to the more detailed nature of emittance compensation in this dc-gun driven injector \cite{serafini1997envelope, Bazarov2005}. In the UED regime, however, the large number of degrees of freedom provides versatility for very high performance both in the stroboscopic (space charge free) regime, and in the single shot (significant space charge) regime.

While having lower gradient, the use of a DC gun does permit significant flexibility in the choice of photocathode and mode of illumination: high quantum efficiency, low emittance photocathodes can be used in both reflection mode and transmission mode (illumination from behind) \cite{Lee:2016transmission}. Transmission mode photocathodes with built-in optical focusing can yield single digit micron initial source sizes \cite{Gerbig2015}. We will demonstrate that in the space charge free regime this can lead to the production of probe sizes below one micron, which would be advantageous to study the dynamics of materials near grain boundaries or domains \cite{Zong2018}. In the single shot regime, we will show that in conjunction with collimating apertures, a novel focusing scheme achieves very high quality emittance compensation for the central core of the beam composing 40\% of particles, for a resulting beam size of 5 $\mu$m (rms), which is well tuned for diffraction from small flakes of advanced materials \cite{Shen2018}.

In the longitudinal dimension, we make use of the fact that MeV UED requires much lower energy than the $15$ MeV max energy of the CBETA injector, and we may therefore use several of the SRF cavities for bunch length compression. In practice, we use a genetic optimization algorithm to choose the phases and amplitudes of the cavities appropriately for optimal bunching. In the zero space charge case, we find that bunching and acceleration is distributed across the 6 cavities in a way that produces a linearizing effect similar to that reported in Ref. \cite{Zeitler:2015fsba}. In this regime, we will show that the ultimate bunch length can be limited by time of flight differences arising from transverse size and transverse momentum spread. 

The minimum bunch lengths achievable in simulation without space charge are well below 1 fs, and in the case of single shot space charge conditions, between 5-20 fs (rms). However, in practice, the ultimate time resolution of the instrument may be limited by time of arrival fluctuations determined by phase and amplitude fluctuations of the accelerating fields. We conclude this work with a parametric study of the sensitivity of the device to these fluctuations.

\section{Instrument layout}

\begin{figure}[!htbp] 
\centering
    \begin{center}
         \subfigure[]{%
            \label{FIG:3d_beamline}
            \includegraphics[width=0.95\columnwidth]{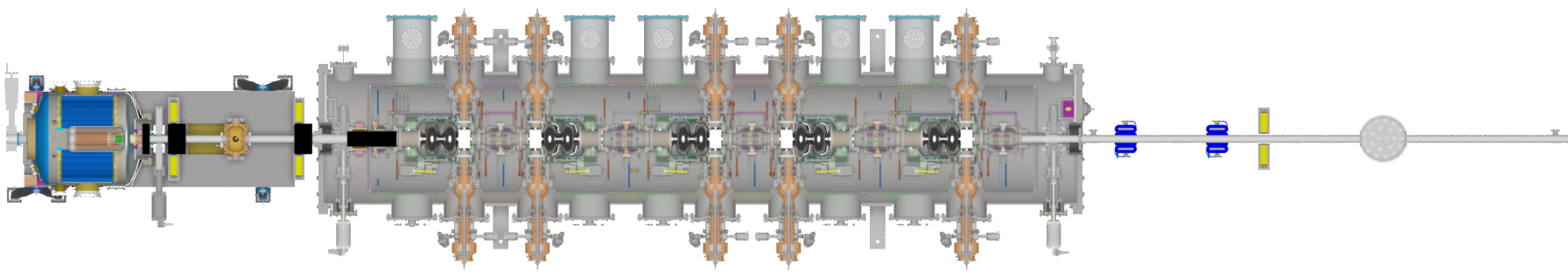}
        }\\
        \subfigure[]{%
           \label{FIG:beamline_fields}
           \includegraphics[width=0.95\columnwidth]{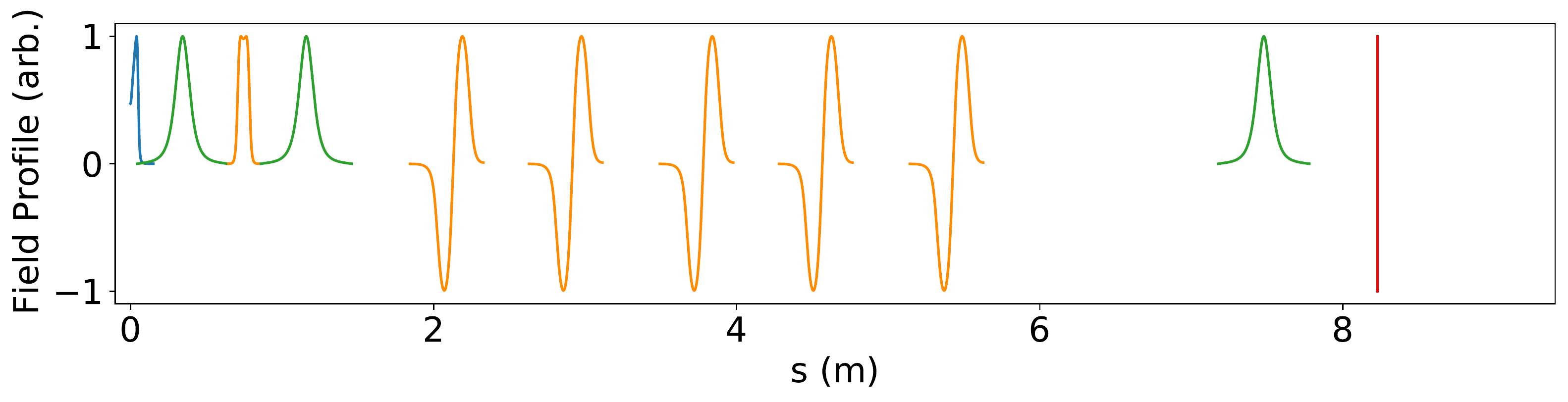}
        }
        \caption{(a) 3D model of the apparatus. (b) Electric and magnetic field profies along the beamline. The DC gun field is shown in blue, solenoid fields in green, RF cavities in orange, and the sample location is shown via a red line.}
        \label{FIG:beamline}
    \end{center}
\end{figure}

The layout of the  injector is shown in Fig. \ref{FIG:beamline} and remains largely unchanged from that in previous works \cite{bazarov2015cornellbnl, ref:lowemitter, doi:10.1063/1.4789395}. It begins with a 400 kV DC gun, followed by a short section that includes two emittance compensating solenoids and a normal conducting RF bunching cavity. Following that is the injector cryomodule (ICM), containing five 2-cell SRF acceleration cavities, each with independent phase and amplitude control, capable of producing a net energy gain up to 15 MeV. As currently installed, the ICM is followed by a four-quadrupole telescope section, which we have modified for this design work.  In the here presented design we replace the final two quadrupole magnets with an additional solenoid and the UED sample chamber. Most of the focusing onto the sample is provided by this solenoid, while any asymmetry in the beam induced by the input couplers to the SRF cavities (which are included in our field model) is cancelled by the remaining two quadrupoles. 

This provides 6 independent RF phase and amplitude knobs, along with 3 solenoid focusing lenses for matching, emittance compensation, and final focusing. The photocathode can be driven by the laser from behind, in transmission mode. This allows for the option to include on-board focusing near the photoemission surface. Analytic estimates and  experimental evidence shows that this can enable initial Gaussian rms beam sizes down to approximately 2 micron \cite{Gerbig2015}. In optimization, we require the initial Gaussian width to be $\sigma_x > 2$ $\mu$m, and we also require that the initial laser pulse length  $\sigma_t > 1$ ps, which is sufficiently long compared to the expected photocathode response time given the nanometer-scale thickness of high quantum efficiency photocathodes \cite{karkare2014ultrabright}. Above those constraints, spot size and bunch length are considered free parameters to be optimized as needed. Final energy is allowed to vary, and all optima shown below lie within 4 and 5 MeV total energy.

\section{Parameter optimization}

In the following section we will limit ourselves to optimization of the beamline’s single-bunch performance, deferring a discussion of the effects of multi-bunch stability until the following section. All of our simulation work is performed with General Particle Tracer (GPT) \cite{ref:gpt1, ref:gpt2}, using a Multi-Objective Genetic Algorithm optimization (MOGA) \cite{ref:lowemitter,ref:lowemitter2,ref:lowemitter3}. Two sets of optimizations were performed, one without space charge which models single electron/pulse conditions, and one with $10^5$ electron/pulse to approximate conditions for single shot diffraction. All optimizations with $10^5$ electron/pulse were performed at a reduced number of simulated macroparticles of 5000, which we found to be sufficient to produce the correct physical trends in the optimized Pareto fronts. After optimizing, select cases were recalculated with 250,000 macroparticles to verify convergence. Typically, the transverse emittance was reduced and the bunch length remained nearly identical after recalculation.

\subsection{Stroboscopic Mode: 1 electron per pulse}
Even when including 3D effects in the fields (aberrations), at zero space charge and with micron-scale initial spot sizes, simulation predicts full emittance preservation, and the emittance is therefore no longer an optimization objective. The emittance is set by initial conditions: the mean transverse energy (MTE) of the photoemission process and the incident laser spot size \cite{Musumeci2018}. Optimizing the machine parameters in this case is much simpler, as there is no need for emittance compensation, and only the bunch's final transverse and longitudinal size need to be optimized. In practice, emittance evolution without space charge effects will be determined by field quality, and is outside the scope of this work. Nonetheless, we see an important trade-off between final spot size and bunch length described below.

Without the need to compensate transverse emittance, cavity parameters may vary freely while maintaining the desired final energy.  This allows advanced longitudinal phase space manipulation techniques to be performed to mitigate the negative effects of nonlinear time of flight and cavity-induced phase space curvature without higher harmonic fields,  similar to what was derived in \cite{Zeitler:2015fsba}.  Extremely short bunch lengths are possible since the phase space can be linearized to high order at the sample plane. We find that the bunch length is then not limited by RF curvature but rather by the coupling of the time of flight to transverse momentum. This coupling of transverse and longitudinal coordinates has been well-studied for sub-femtosecond beams  \cite{de2006radial, Floettmann2014, Weninger2012}, and here we find that it is this phenomenon which sets the trade-off between ultimate bunch length and spot size.

Because our beam begins with a tight focus at the cathode, the downstream trajectory of each emitted electron is determined almost entirely by its initial transverse momentum. Larger initial momenta will produce trajectories that deviate further from the transverse center, causing a longer path length to the sample. Thus, for a fixed set of machine optics in this apparatus, the minimal bunch length in the sample plane is primarily a function of the cathode's MTE, and depends only weakly on the initial transverse or longitudinal size.

The final transverse beam size is determined by the emittance of the beam and by how strongly the beam is focused onto the sample. As the emittance increases, achieving a given final beam width will require stronger focusing, which is only possible with larger beam widths at the focusing optics. But, larger widths during beam transport also produces larger spreads in time-of-flight. That is, for a given final beam width, a larger emittance will result in a longer bunch length. This is opposite to the typical behavior of space charge dominated beams. In addition, for a constant emittance, there will be a trade-off between beam size and bunch length. 

\begin{figure}[htp]
	\centering
		\includegraphics[width=0.9\columnwidth]{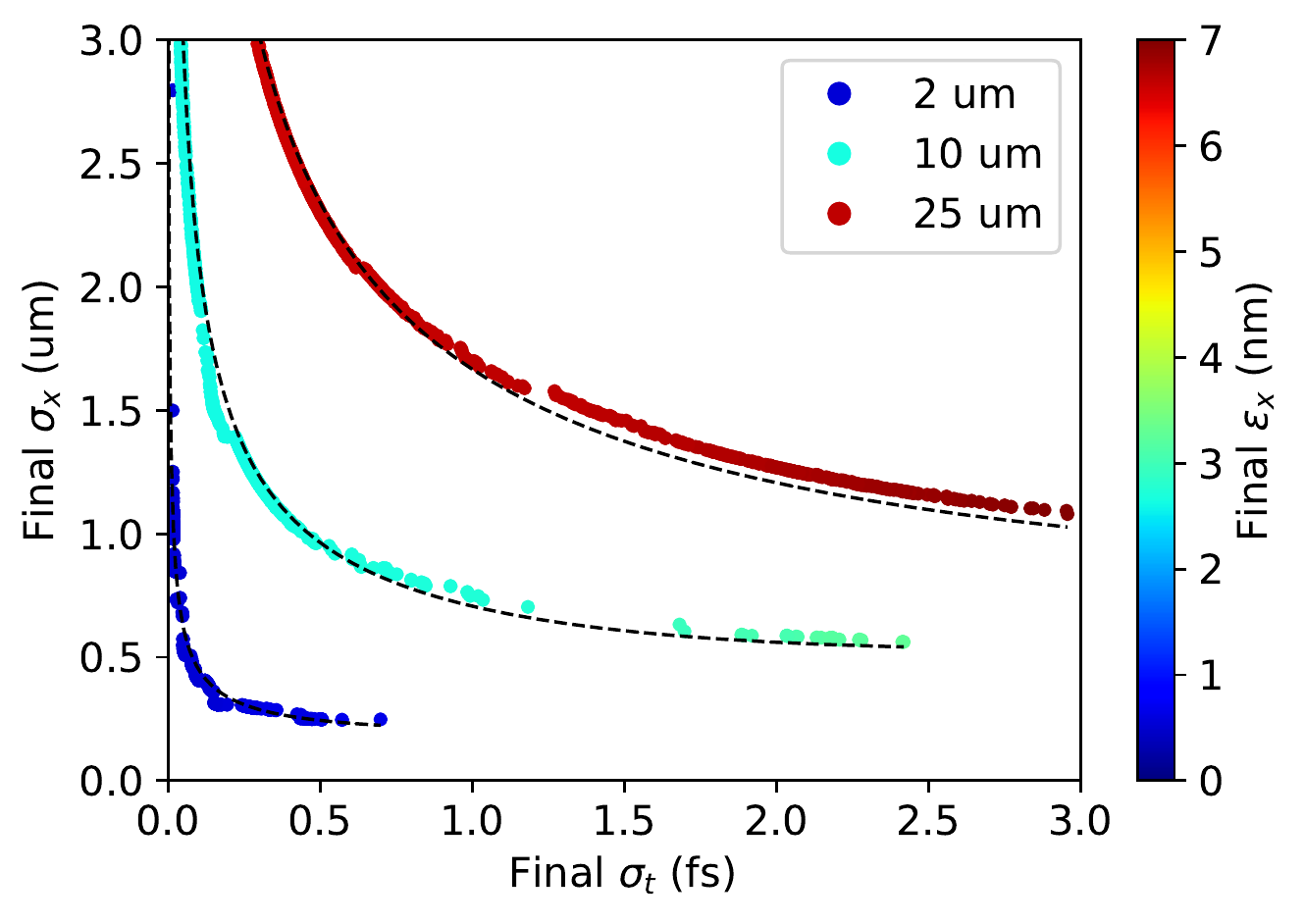}
	\caption{Dependence of achievable bunch length on final beam size in the absence of space charge forces for three different initial laser sizes: 2, 10, and 25 $\mu$m (dots), compared to a simple model (lines). Here the MTE is 35 meV, and so these laser sizes correspond to normalized source emittances of 0.5 nm, 2.6 nm, and 6.5 nm, respectively.  }
	\label{FIG:no_SC_with_model}
\end{figure}

We can get insight into both of these trends by looking at the spread in path lengths near the sample. At the sample location, if the emittance and beam size are known, then the momentum spread can be directly determined as $\sigma_{p_x}= m c \epsilon_n/\sigma_x$. For each individual electron, the time of flight from the last focusing magnet to the sample is $\Delta t_f = L/v_z = L m\gamma / \sqrt{p^2-p_x^2-p_y^2}\approx L \frac{m\gamma}{p}(1+\frac{1}{2})\frac{p_x^2+p_y^2}{p^2}$, with an rms spread of
$\sigma_{t_f} = L \frac{m\gamma}{\sqrt{2}p^3}\sqrt{<p_x^4> - \sigma_{p_x}^4}$.
Assuming a Gaussian spread in momenta, this produces a spread in time-of-flights back to the location of the final focusing magnet given by:
\begin{equation}
\sigma_{t_f} = \frac{L \epsilon_n^2}{\beta c (\gamma \beta)^2 \sigma_x^2}
\end{equation}

where L is the drift length from the lens to the sample, $\epsilon_n$ is the normalized beam emittance, and $\sigma_x$ is the beam size at the sample. This only describes the spread in time-of-flights induced by the last focusing magnet. Because the final focus produces the largest transverse momentum spread, this section dominates the time of flight spread. Fig. \ref{FIG:no_SC_with_model} shows the optimal Pareto fronts from the MOGA optimizations for three different choices of initial laser size, along with a fit to this model using a scaling constant of $2$. That is, the contribution to the spread in path lengths before the last focusing magnet is proportional to that after the magnet for those optics that are located on the Pareto front. This proportionality can be understood as follows: To achieve a small $\sigma_x$ for a specified emittance, a large beamsize $\sigma_{x_f}\propto 1/\sigma_x$ at the final focusing element is required. The time of flight spread $\sigma_{t_f}$ resulting from the large $\sigma_{p_x}$ that is needed to focus this large size down to $\sigma_x$ is unavoidable and is similar in all optics settings, not only for the one on the Pareto front. The optics with the smallest bunch length at the sample is the one that magnifies the beam size on the cathode to the large $\sigma_{x_f}$ in the smoothest possible way, producing the smallest average $\sigma_{p_x}^2$. For all these solutions, the average $\sigma_{p_x}^2$ is roughly proportional to $\sigma_{x_f}^2$ and therefore also proportional to $\sigma_{t_f}$. 

In all simulations, the cathode MTE was set to 35 meV, assuming near threshold illumination of alkali antimonide photocathodes, which have been the standard for this electron gun \cite{Maxson2015}. 

\begin{figure}[!htbp] 
\centering
    \begin{center}
         \subfigure[]{%
            \label{FIG:tpz_long}
            \includegraphics[width=0.9\columnwidth]{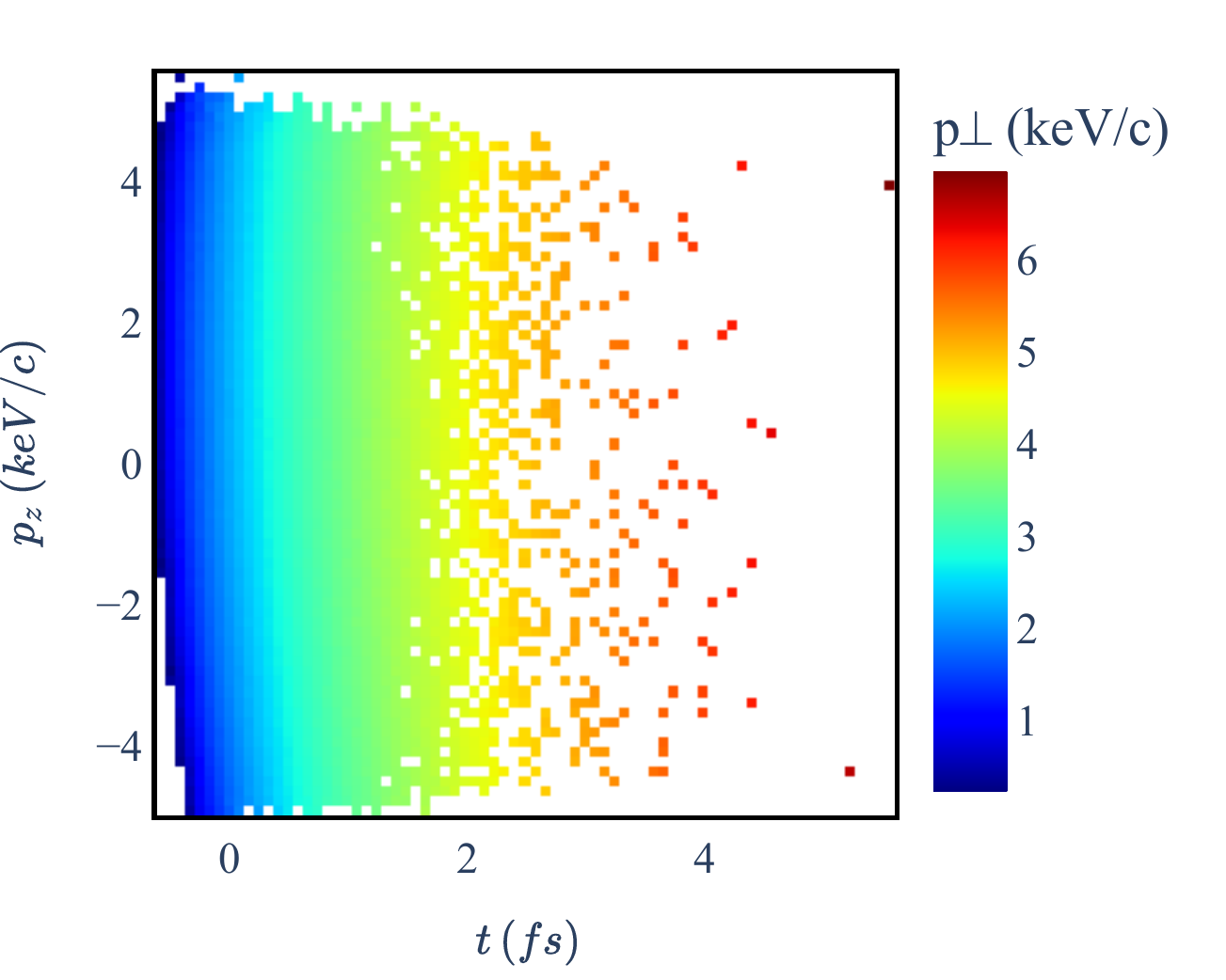}
        }\\
        \subfigure[]{%
           \label{FIG:tpz_short}
           \includegraphics[width=0.9\columnwidth]{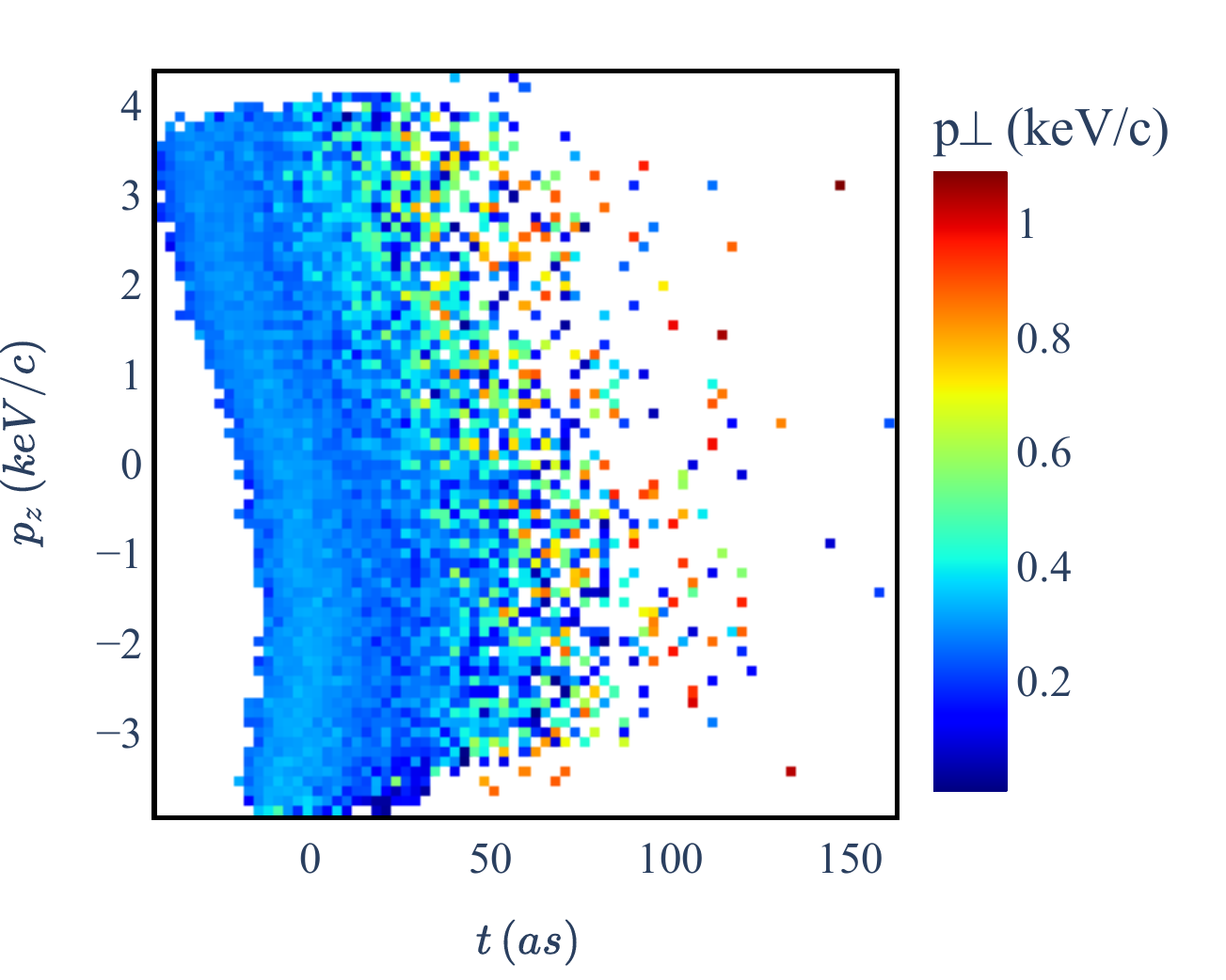}
        }
        \caption{Longitudinal phase spaces of an optimized bunch at the sample location at low charge. Examples are shown for both (a) typical and (b) minimal bunch lengths. Particles are colored according to their transverse momentum, showing that this is the limiting effect for (a), but does not contribute significantly in (b).}
        \label{FIG:no_SC_long_dist}
    \end{center}
\end{figure}

\begin{figure}[!htbp] 
\centering
    \begin{center}
         \subfigure[]{%
            \label{FIG:trend_beam_size_noSC}}
            \includegraphics[width=0.9\columnwidth]{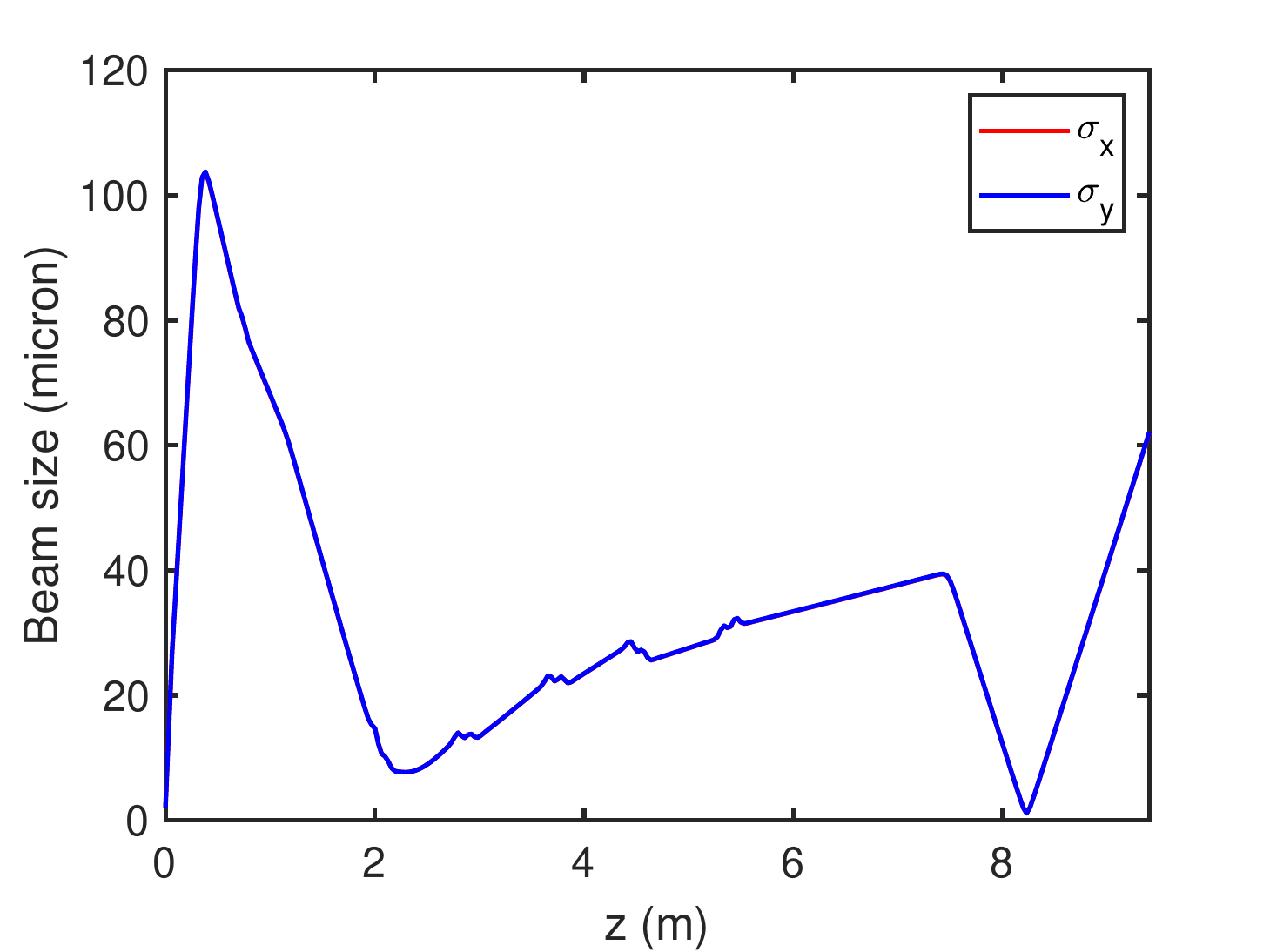}
        \\
        \subfigure[]{%
           \label{FIG:trend_bunch_length_noSC}
           \includegraphics[width=0.9\columnwidth]{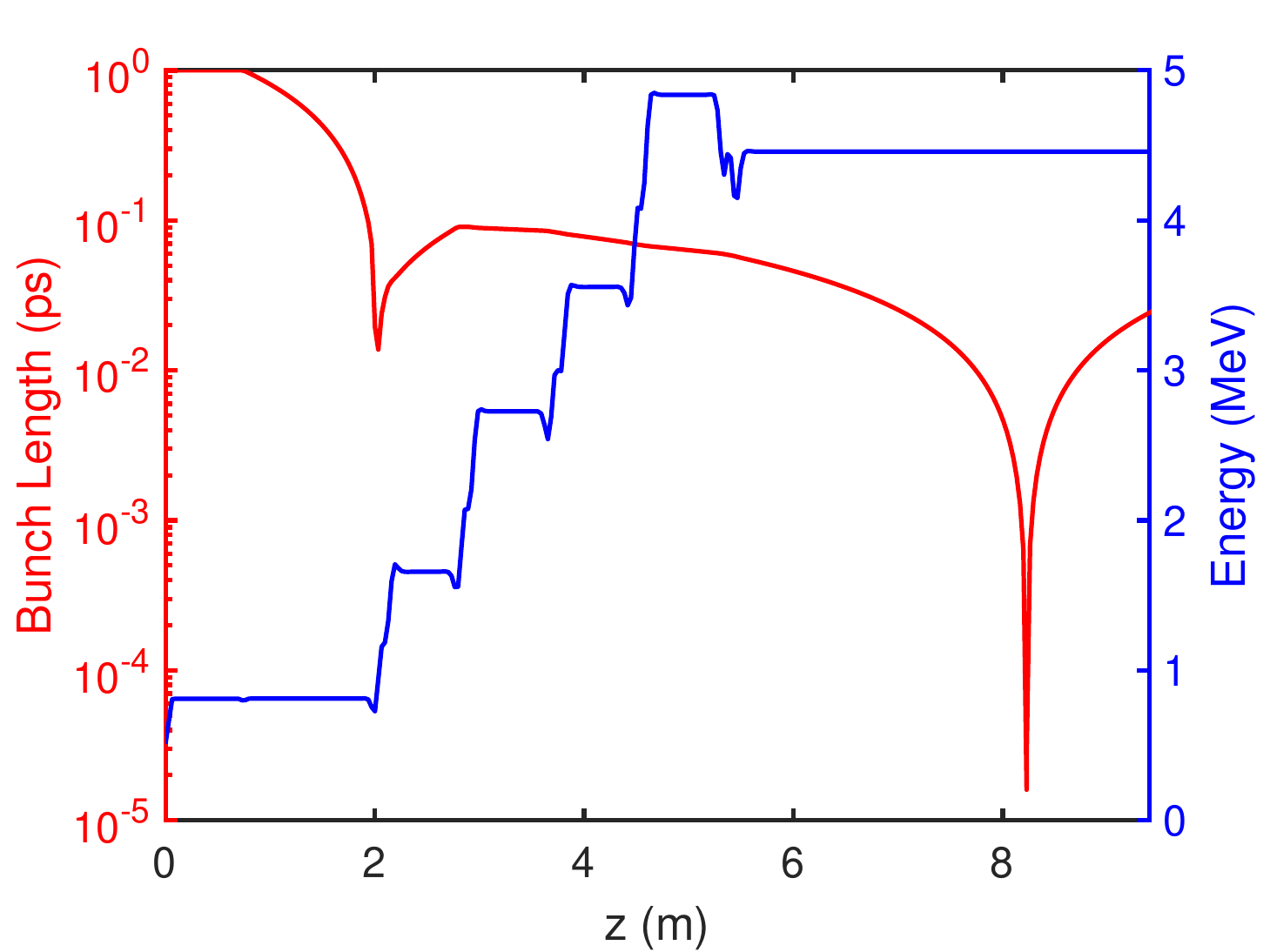}
        }
        \caption{(a) Beam width and (b) bunch length and energy throughout the injector at near zero charge for an example solution on the pareto front. The sample (and clipping aperture) is located at $z=8.228$ m. }
        \label{FIG:trends_without_space_charge}
    \end{center}
\end{figure}

To illustrate this effect, two example longitudinal phase spaces are shown in Fig. \ref{FIG:no_SC_long_dist}. For most of the optimal points on the front, the beam is similar to Fig \ref{FIG:tpz_long}, where there is a noticeable correlation between final arrival time and transverse momentum. But for the smallest bunch lengths, where $\sigma_{p_x}$ at the sample is small, this correlation is absent in the final distribution as shown in Fig. \ref{FIG:tpz_short}. Instead, two other effects are noticeable. First, there is a remaining cubic dependence on $p_z$, from incomplete cancellation of RF effects, though this is not dominant. There is also an overall spread in arrival time that is due to the initial longitudinal energy spread on the cathode, and better cathodes with improved momentum spread would be able to push these bunch lengths even lower. For one example on the Pareto front, we plot the transverse beam size, bunch length, and energy in Fig. \ref{FIG:trends_without_space_charge}.

It must be noted that the final achievable bunch lengths are in the range of a few femtoseconds down to $<100$ attoseconds (rms), which suggests multi-cavity injectors of this type as a promising route to generate high repetition rate, isolated attosecond electron pulses. However, in practice with the current state of the art, we will show below that UED temporal resolution will be limited by the phase and amplitude stability of the accelerating fields; time-stamping techniques \cite{zhao2021noninvasive} with sufficient precision would be needed to overcome this limitation.

\subsection{Single shot mode: $10^5$ electrons per pulse}

At a charge of 16 fC, with $10^5$ electrons, space charge becomes the dominant effect in transverse and longitudinal dynamics and careful emittance compensation is required.  Optimizations were performed to determine to what extent these effects can be mitigated in this apparatus, which has much smaller source accelerating gradient ($\sim 5$ MV/m) than other MeV UED machines which are based on RF guns. MOGA optimizations were started with three objectives: beam size, emittance, and bunch length. Once roughly converged, the beam size objective was replaced with a constraint ($\sigma_x < 30$ $\mu$m), and the MOGA was continued with only 2 objectives. This beam size constraint was chosen so that even with significant space charge, diffraction from most samples can be achieved. In practice, the optimizer would tend to aggressively focus the beam, and was typically well under the 30 $\mu$m beam size constraint. This is atypical of space charge optimizations, and we believe it was due to our use of an adjustable aperture, as explained below.

An aperture directly before the sample location is a practical way to force the size of the electron beam to be compatible with a given sample size. In addition, if one emits more charge than is necessary, the final sample pinhole can be used to improve the transverse emittance performance of the device by selecting only the beam's dense core, which has experienced a more linear space charge force overall. To exploit this, we allowed the optimization to begin with a larger charge of 40 fC, and then to subsequently clip back to 16 fC at the sample location with an adjustable circular aperture. We chose that value of initial charge because we found that increasing the initial charge above 40 fC produced only marginal improvement, while requiring longer simulation times. In each simulation, the radius of the aperture was chosen to produce precisely the target charge of 16 fC, with typical aperture sizes $\sim10$ $\mu$m diameter. Apertures of this size appropriate for few-MeV level electron beams can be manufactured via laser machining of thin metal films. The optimizer found that the effect of strong focusing would be to produce a bright central beam core, surrounded by a large diffuse halo. Without an aperture, this halo would ruin the beam's emittance and brightness, but with an aperture it was an overall improvement. As a result, the optimizer would choose to focus strongly on small apertures. 

\begin{figure}[!htbp] 
\centering
    \begin{center}
         \subfigure[]{%
            \label{FIG:collimation}
            \includegraphics[width=0.9\columnwidth]{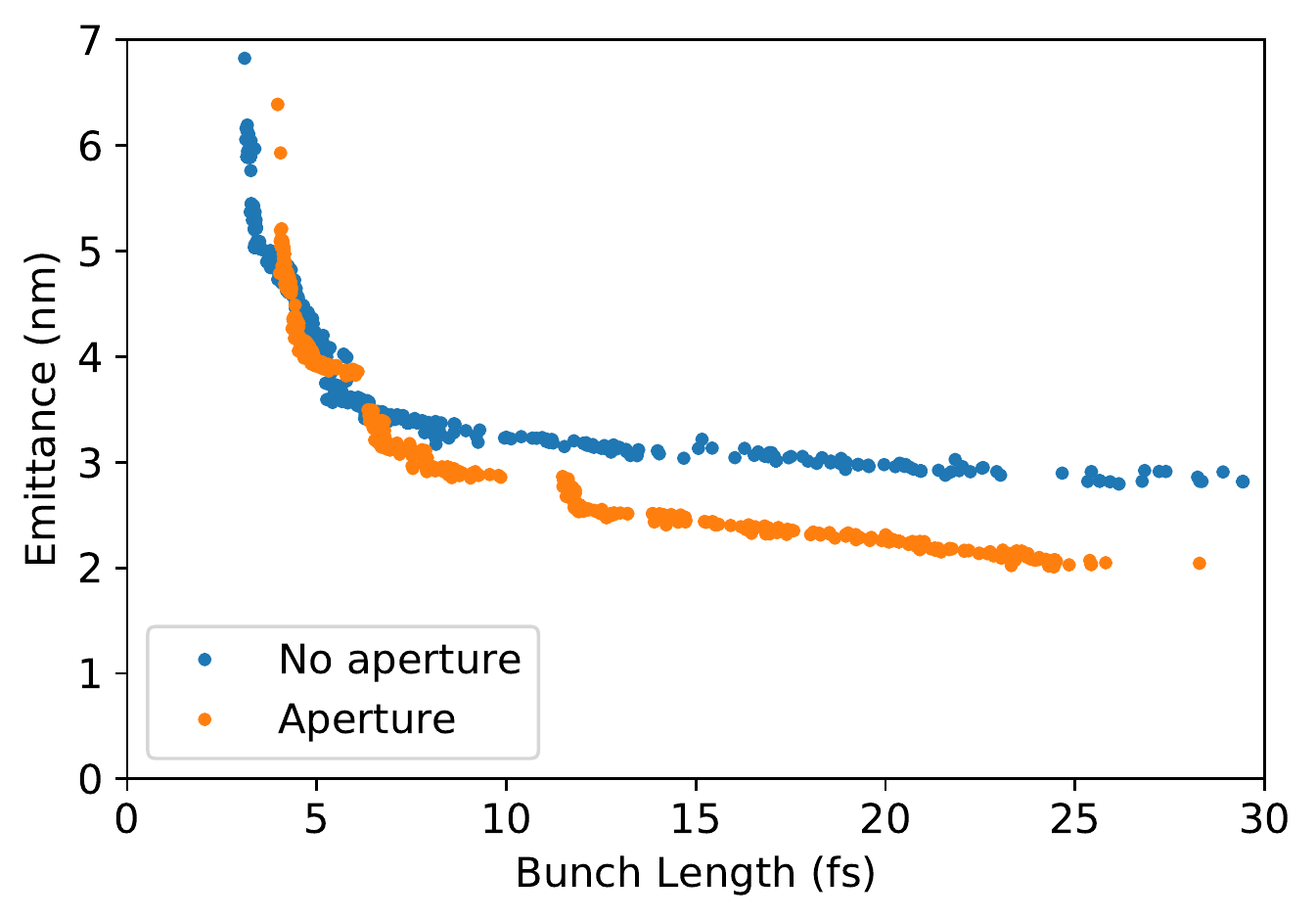}
        }\\
        \subfigure[]{%
           \label{FIG:collimation_MTE}
           \includegraphics[width=0.9\columnwidth]{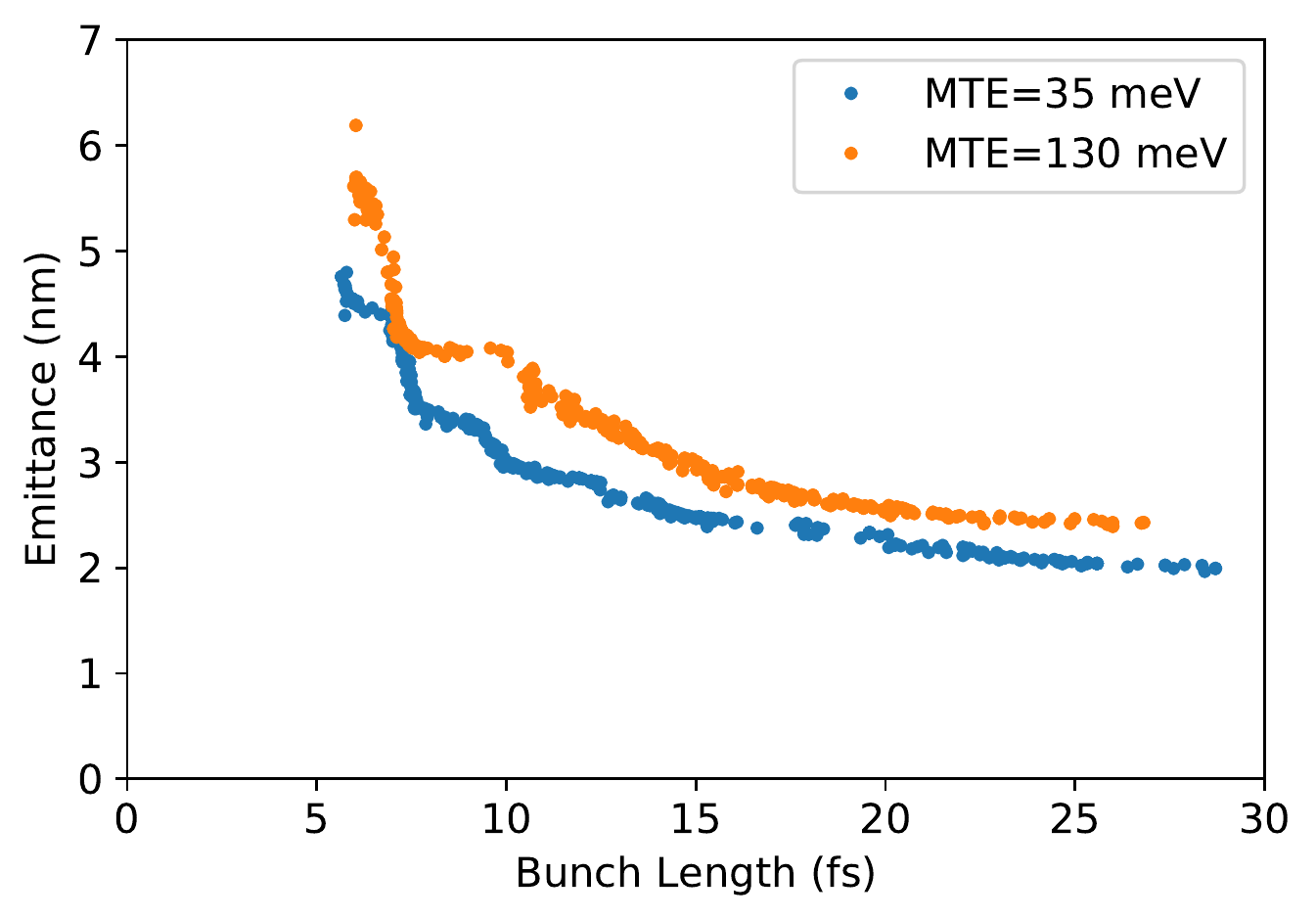}
        }
        \caption{(a) Influence of collimation on the achievable bunch length and emittance with a final bunch charge of 16 fC. (b) Effects of the cathode mean transverse energy (MTE) on the final bunch length and emittance at the same charge. }
        \label{FIG:optimization_fronts}
    \end{center}
\end{figure}

The optimization results with and without using a sample-plane aperture are shown in Fig. \ref{FIG:collimation}. For most bunch lengths, the emittance is significantly reduced by the aperture, although the minimum achievable bunch length is slightly worse due to the requirement of starting with a larger initial charge, and due to transverse/longitudinal spatial correlations present in the bunch prior to the aperture. To determine the impact of the photoemission MTE on the final properties of the beam, simulations were performed at both an MTE of 35 meV and 130 meV, as shown in Fig. \ref{FIG:collimation_MTE}. The latter is typical for multi-alkali cathodes when illuminated with green light, which is often easier to obtain via doubling of commercially available Yb-based high repetition rate lasers, while the former requires a laser tuned to near-threshold illumination. Experimentally reaching 35 meV at the particle densities simulated here is challenging due to the onset of multi-photon photoemission. If a significant fraction of the bunch's electrons are excited from more than a single photon, there would be two populations of electrons: those with low MTE (singly excited electrons) and those with high MTE (doubly excited electrons). In momentum space, this distribution will look like a sharply peaked Gaussian atop a wider one. Therefore, even if this effect becomes important, it may be possible to eliminate the contribution of multi-photon photoemission via apertures. Regardless, it is valuable to model the effect of lower MTE to judge the potential for improvement.

\begin{figure}[!htbp] 
\centering
    \begin{center}
         \subfigure[]{%
            \label{FIG:trend_beam_size}
            \includegraphics[width=0.9\columnwidth]{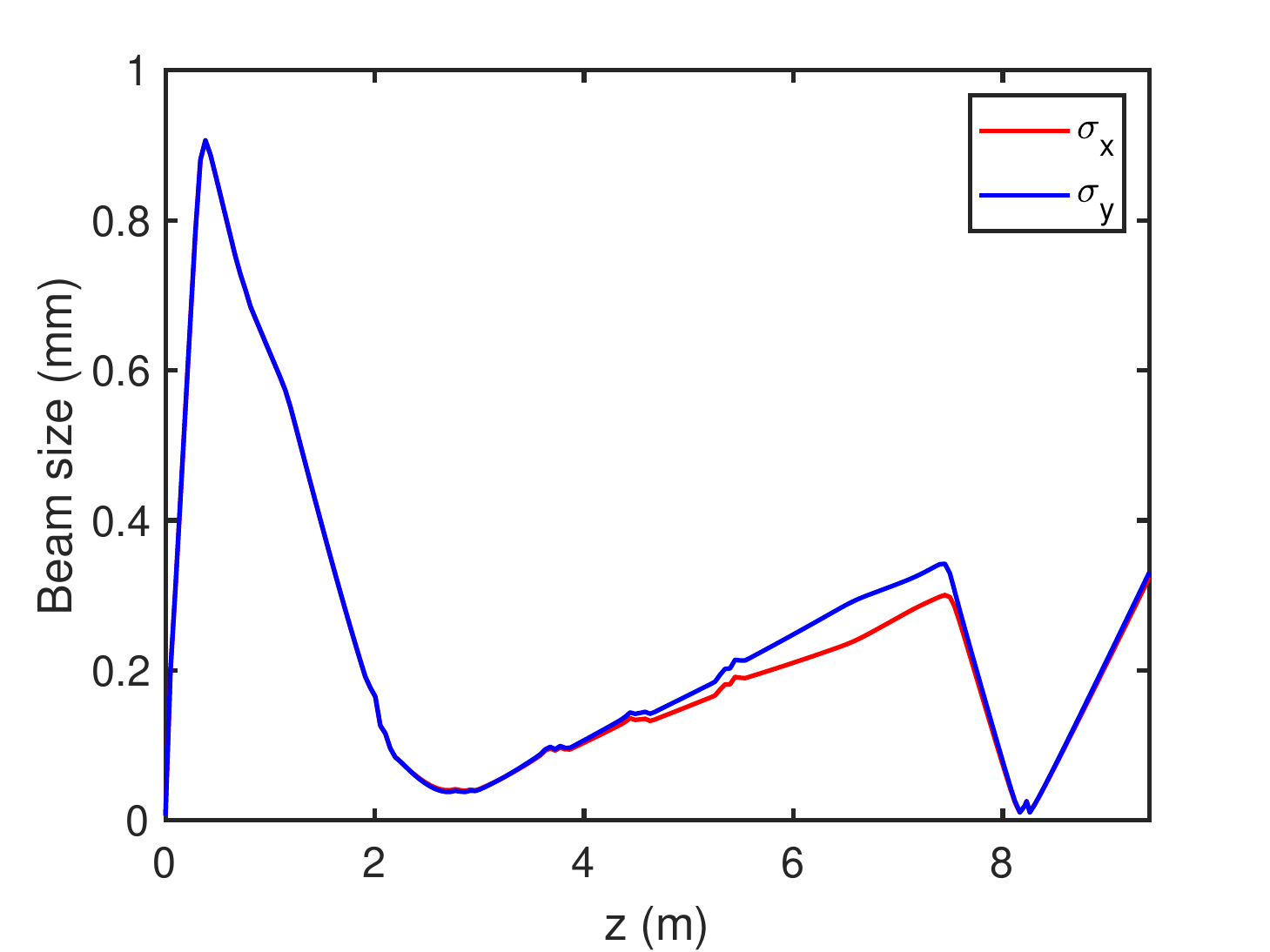}
        }\\
        \subfigure[]{%
           \label{FIG:trend_bunch_length}
           \includegraphics[width=0.9\columnwidth]{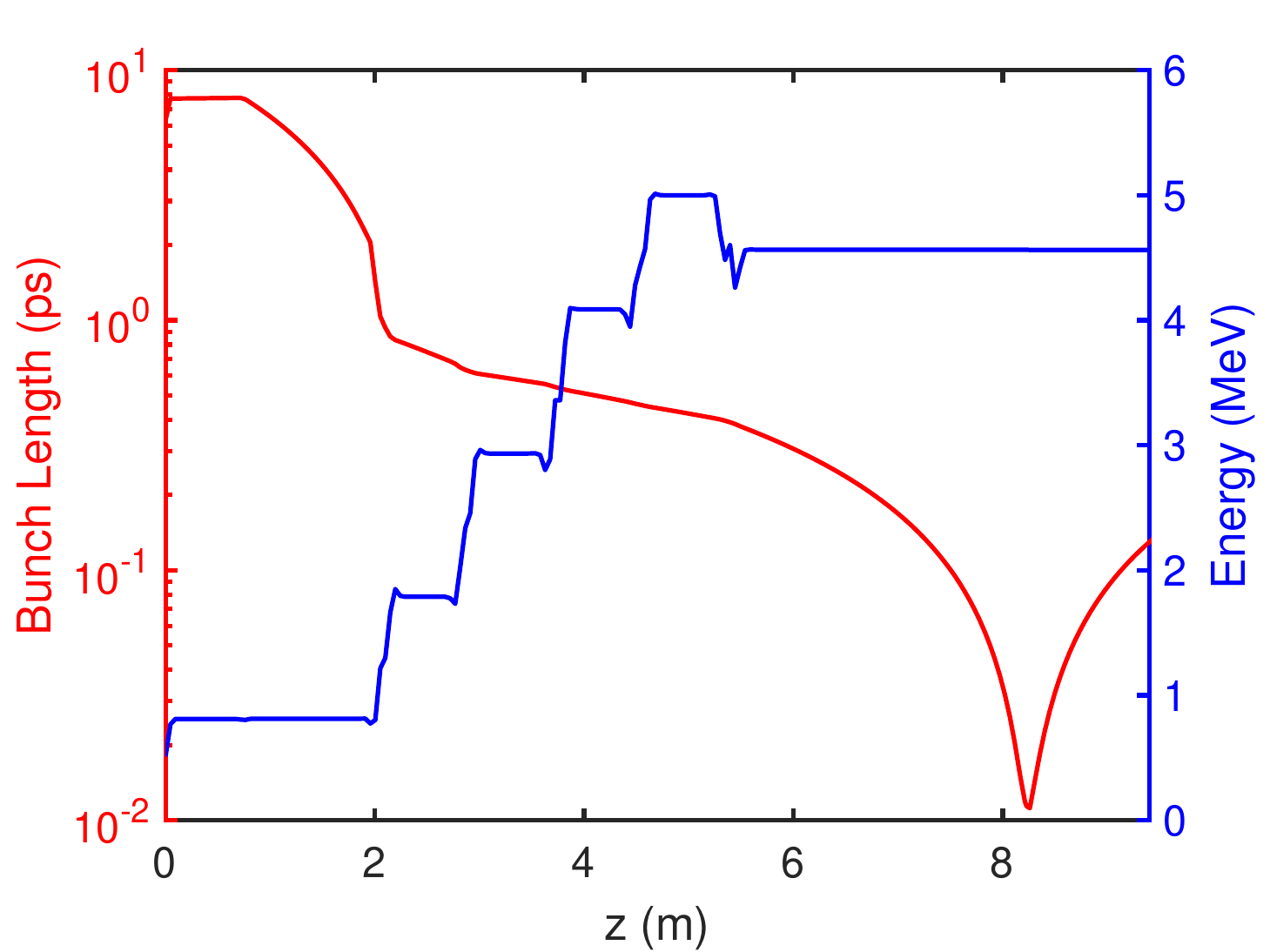}
        }
        \caption{(a) Beam width and (b) bunch length and energy throughout the injector at a final charge of $10^5$ electrons/pulse. After the sample (and clipping) aperture at $z=8.228$ m the bunch had an emittance of 2.6 nm, beam width of 5 $\mu$m, and bunch length of 10 fs.}
        \label{FIG:trends_with_space_charge}
    \end{center}
\end{figure}

\begin{figure*}[!htbp] 
\centering
    \begin{center}
         \subfigure[Transverse distribution before aperture]{%
            \label{FIG:xy_before}
            \includegraphics[width=0.9\columnwidth]{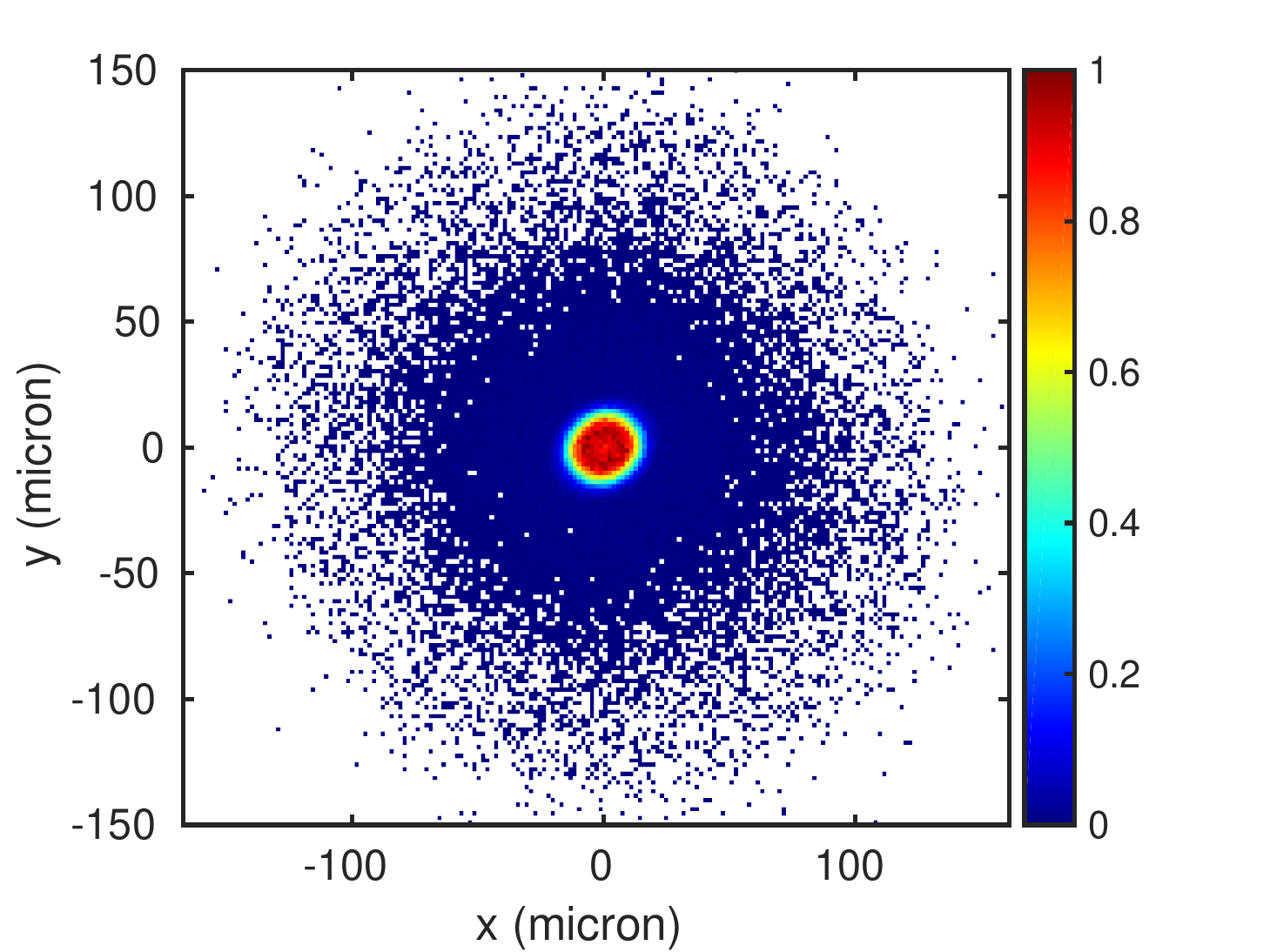}
        }
        \subfigure[Transverse distribution after aperture]{%
           \label{FIG:xy_after}
           \includegraphics[width=0.9\columnwidth]{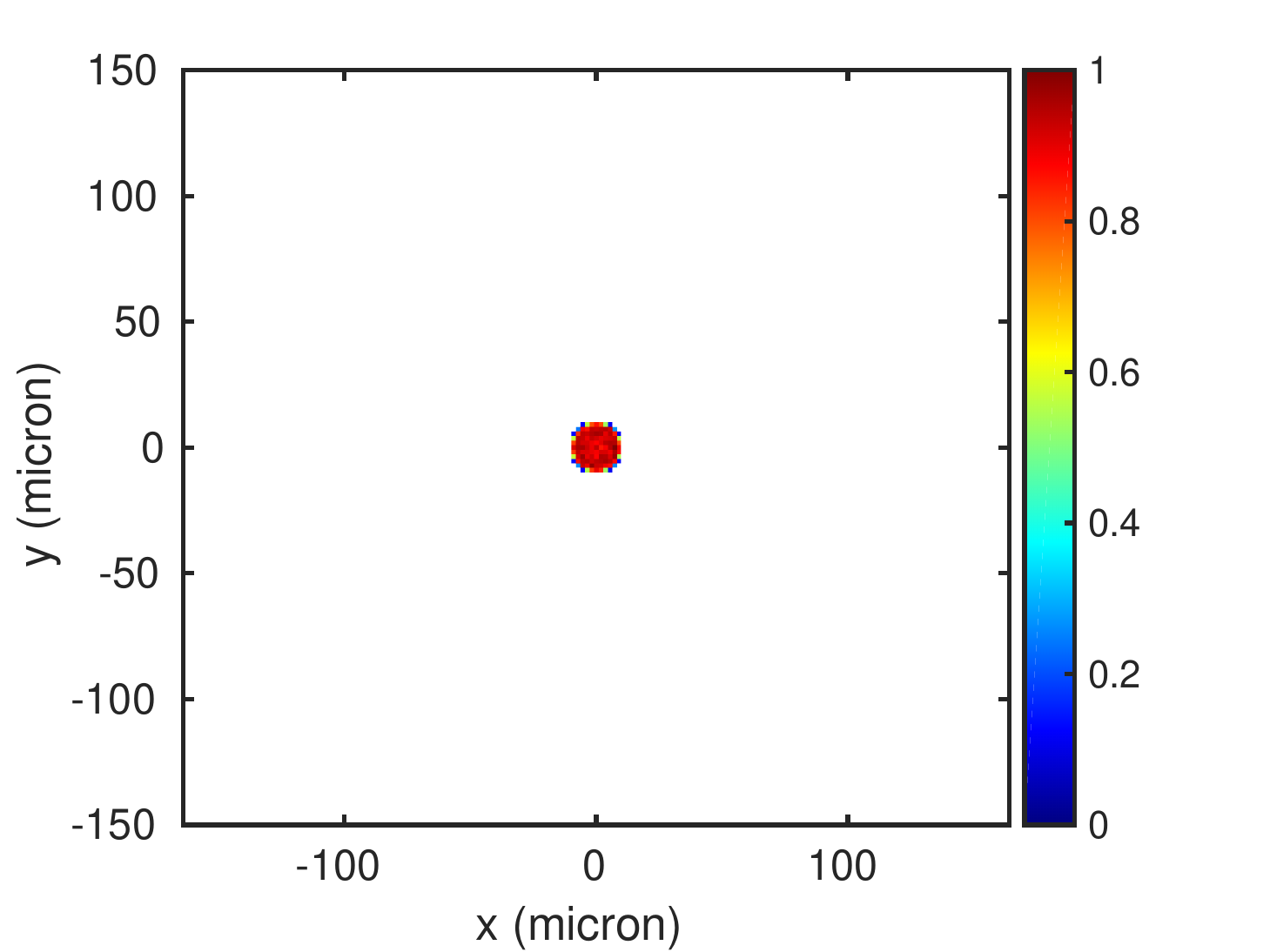}
        }\\
        \subfigure[Longitudinal phase space before aperture]{%
           \label{FIG:tpz_before}
           \includegraphics[width=0.9\columnwidth]{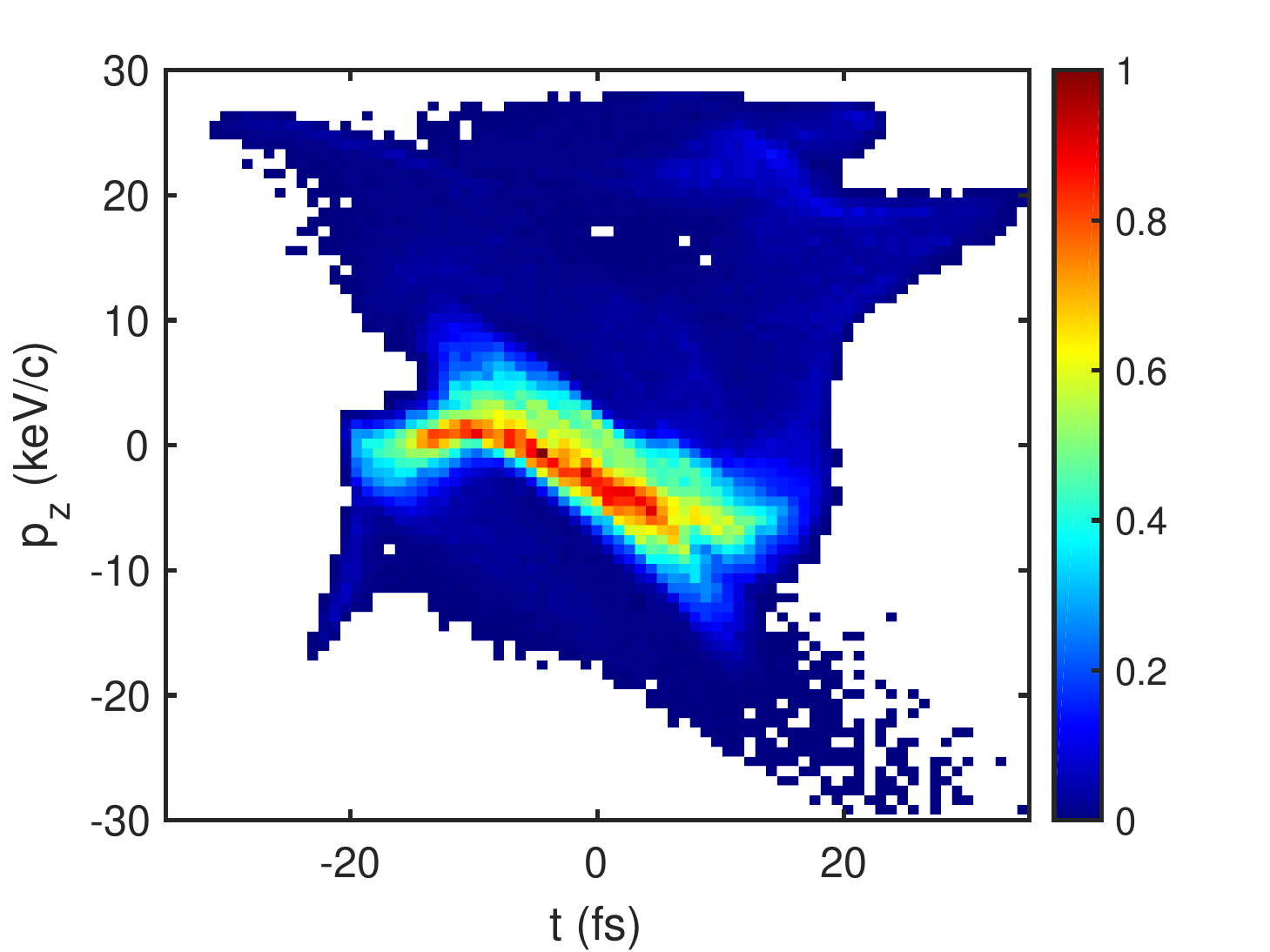}
        }
        \subfigure[Longitudinal phase space after aperture]{%
           \label{FIG:tpz_after}
           \includegraphics[width=0.9\columnwidth]{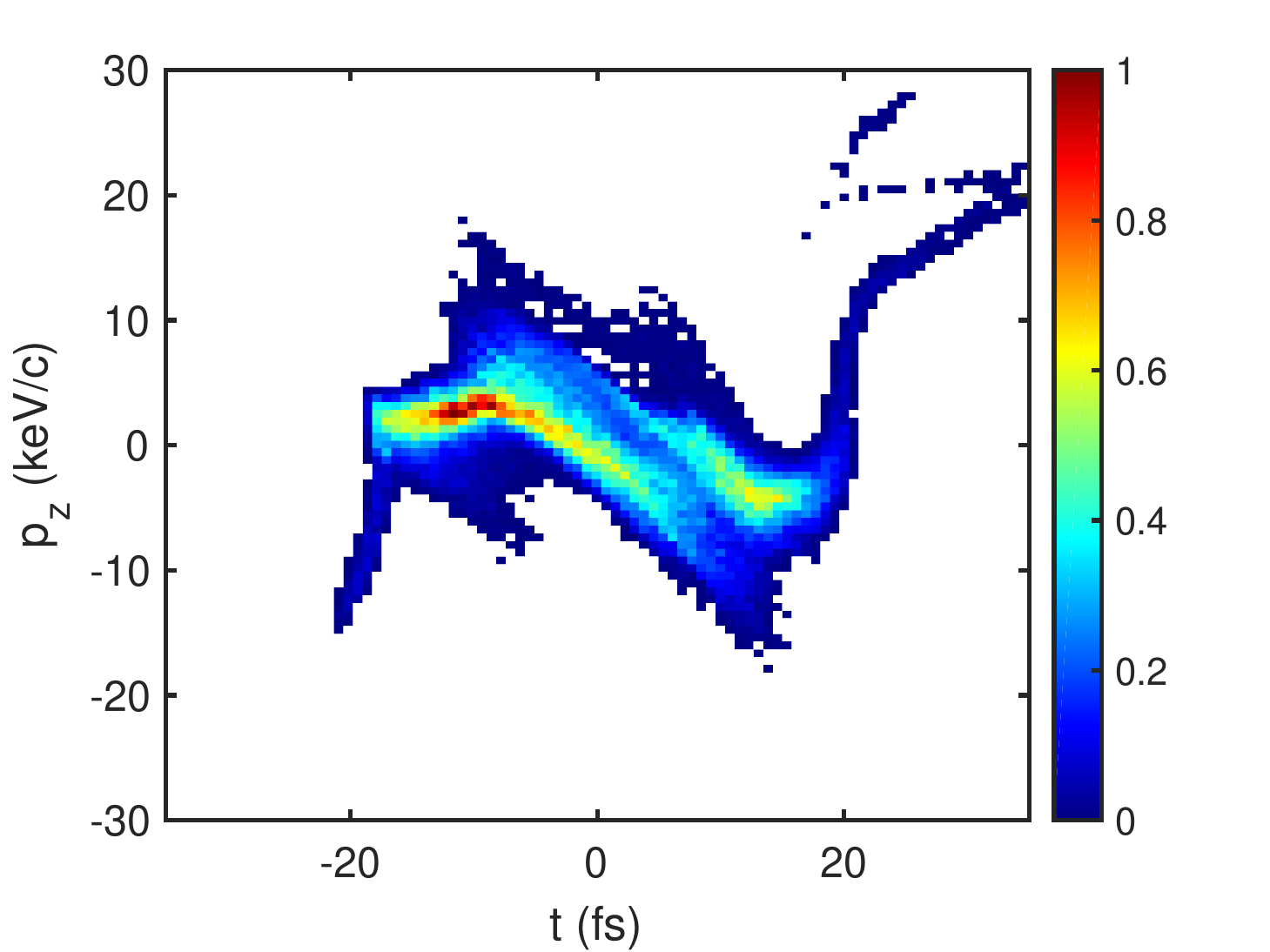}
        }
        \caption{Bunch transverse distribution and longitudinal phase space before and after going through the aperture at the sample location at a final charge of $10^5$ electrons/pulse. After the aperture, the bunch had an emittance of 2.6 nm, beam width of 5 $\mu$m, and bunch length of 10 fs.}
        \label{FIG:phase_spaces}
    \end{center}
\end{figure*}

\begin{figure}[htp]
	\centering
		\includegraphics[width=0.9\columnwidth]{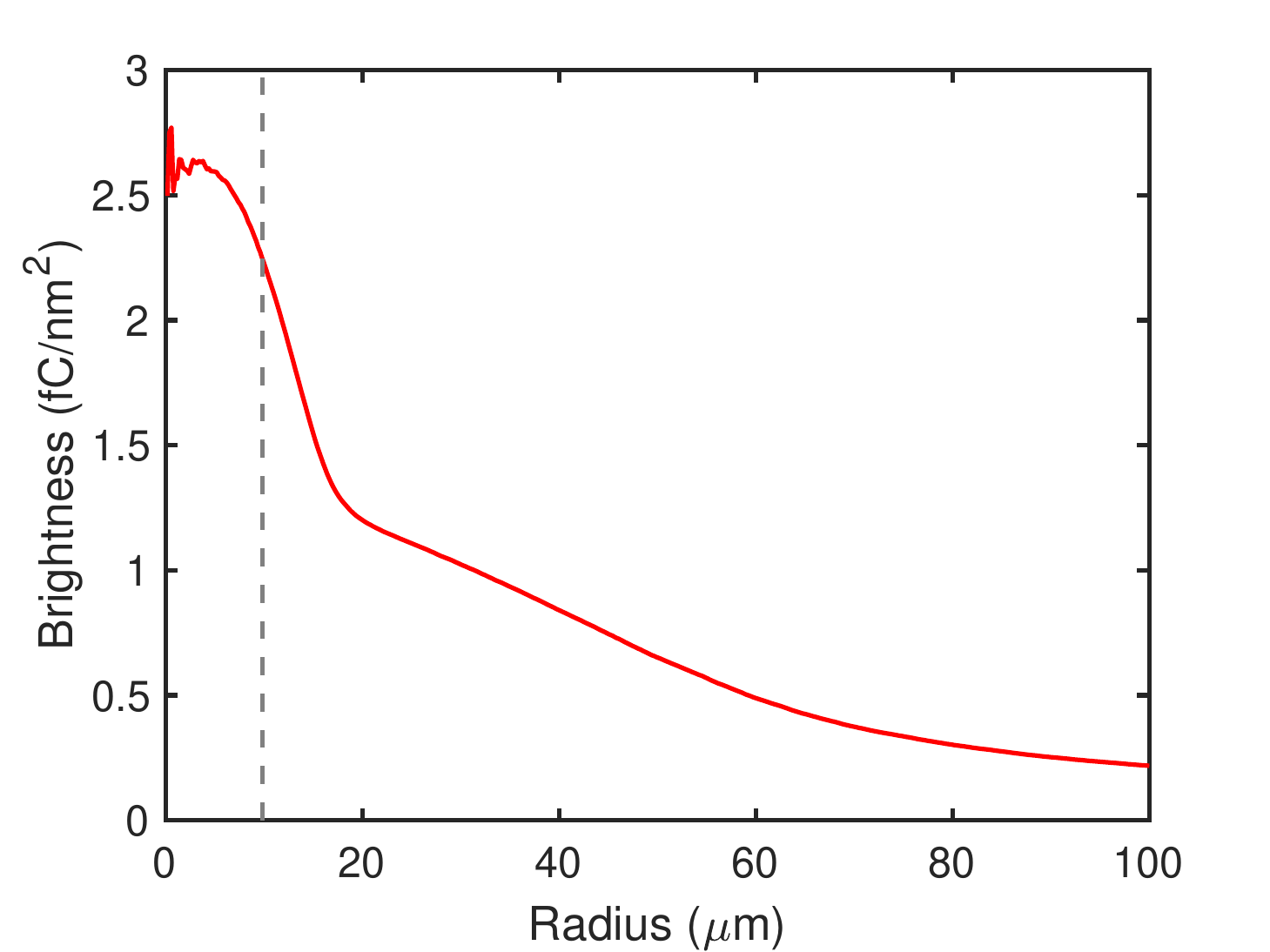}
	\caption{Dependence of transverse brightness on the clipping aperture's radius. The dashed line indicates the radius used in optimization.}
	\label{FIG:brightness_trend}
\end{figure}

To look more closely at the dynamics of the bunch, an example was chosen from the 130 meV MTE Pareto front, and re-evaluated at higher simulation accuracy settings. As is typical for space charge simulations, the higher accuracy slightly reduces the final beam emittance from the value shown in the Pareto front, so that an emittance of 3 nm is achievable at a bunch length of 10 fs. Beam width and bunch length throughout the beamline are shown in Fig. \ref{FIG:trends_with_space_charge}, and example particle distributions are shown in Fig. \ref{FIG:phase_spaces} both before and after the final clipping aperture. Because of the correlation between transverse position and arrival time, clipping occurs in both transverse and longitudinal planes. The aperture improves not only the emittance, but also the transverse brightness of the electron bunch $B=Q/\epsilon_x \epsilon_y$. In Fig. \ref{FIG:brightness_trend}, the radius of the aperture is varied and the brightness of the remaining particles is shown to approach a constant value when only the core of the beam remains. The space charge dynamics of this focusing scheme is more general than this application alone and justifies a detailed analysis in future work. We note that the 5d brightness simulated, with bunch length and emittance in the range of 10 fs and 2-3 nm (depending on MTE), exceeds the experimental state of the art \cite{maxson2016direct, Shen2018, denham2021analytical}, and naturally permits operation at very high repetition rate.

\section{Time of arrival variation due to phase and amplitude stability}

While the design choice of many independent cavity phases and amplitudes provides several advantages, for a given achievable phase and amplitude stability, more cavities can lead to worse time of arrival stability. Without time-stamping techniques to determine post-facto the arrival time of the bunch, this time of arrival variation directly translates to reduced temporal resolution. We define a standard figure of merit for the temporal resolution of any measurement which is a combination of the bunch length $\sigma_t$ and the arrival time stability $\sigma_{t_0}$: $\sigma_{t, \text{res}} = \sqrt{\sigma_t^2 + \sigma_{t_0}^2}$. Because of the large longitudinal compression of our beamline, the effect of pump laser stability will be nearly completely uncorrelated to the e-beam, and simply add in quadrature to bunch arrival stability. The methods used to reduce laser arrival jitter are beyond the scope of this work, and will not be discussed. Instead, we restrict ourselves to the arrival jitter of the electron bunches. 

To determine $\sigma_{t_0}$, we use previously experimentally measured stability of each machine component \cite{Liepe:2010jza} and performed many simulations with settings chosen randomly from the measured distributions. In addition, we also performed simulations that only allowed a single machine parameter to vary, in order to see which parameters individually mattered the most. Using the same example machine setting as in Figs. \ref{FIG:trends_with_space_charge}-\ref{FIG:brightness_trend}, Table \ref{tab:machine_settings} shows the measured stability of each machine parameter, and the individual contribution of it towards arrival time stability. The achievable arrival time stability is around 13 fs, with the most significant contributions coming from the the phase of the off-crest cavities, which is to be expected. As a result, the bunch length and arrival time stability will contribute nearly equally to the total temporal resolution in this case.

\begin{figure}[!htbp] 
\centering
    \begin{center}
         \subfigure[]{%
            \label{FIG:stability_fronts}
            \includegraphics[width=0.85\columnwidth]{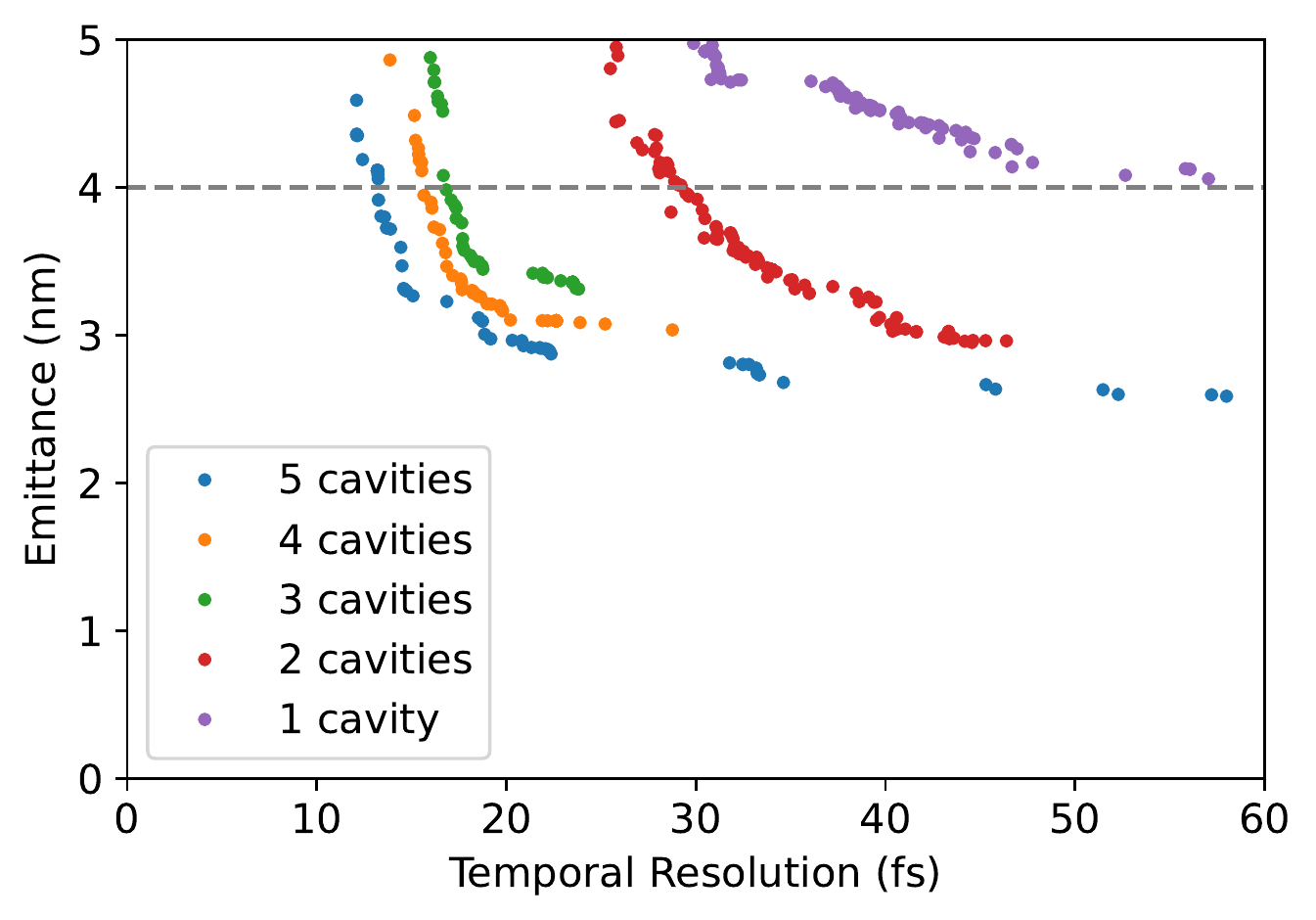}
        }\\
        \subfigure[]{%
           \label{FIG:stability_lines}
           \includegraphics[width=0.90\columnwidth]{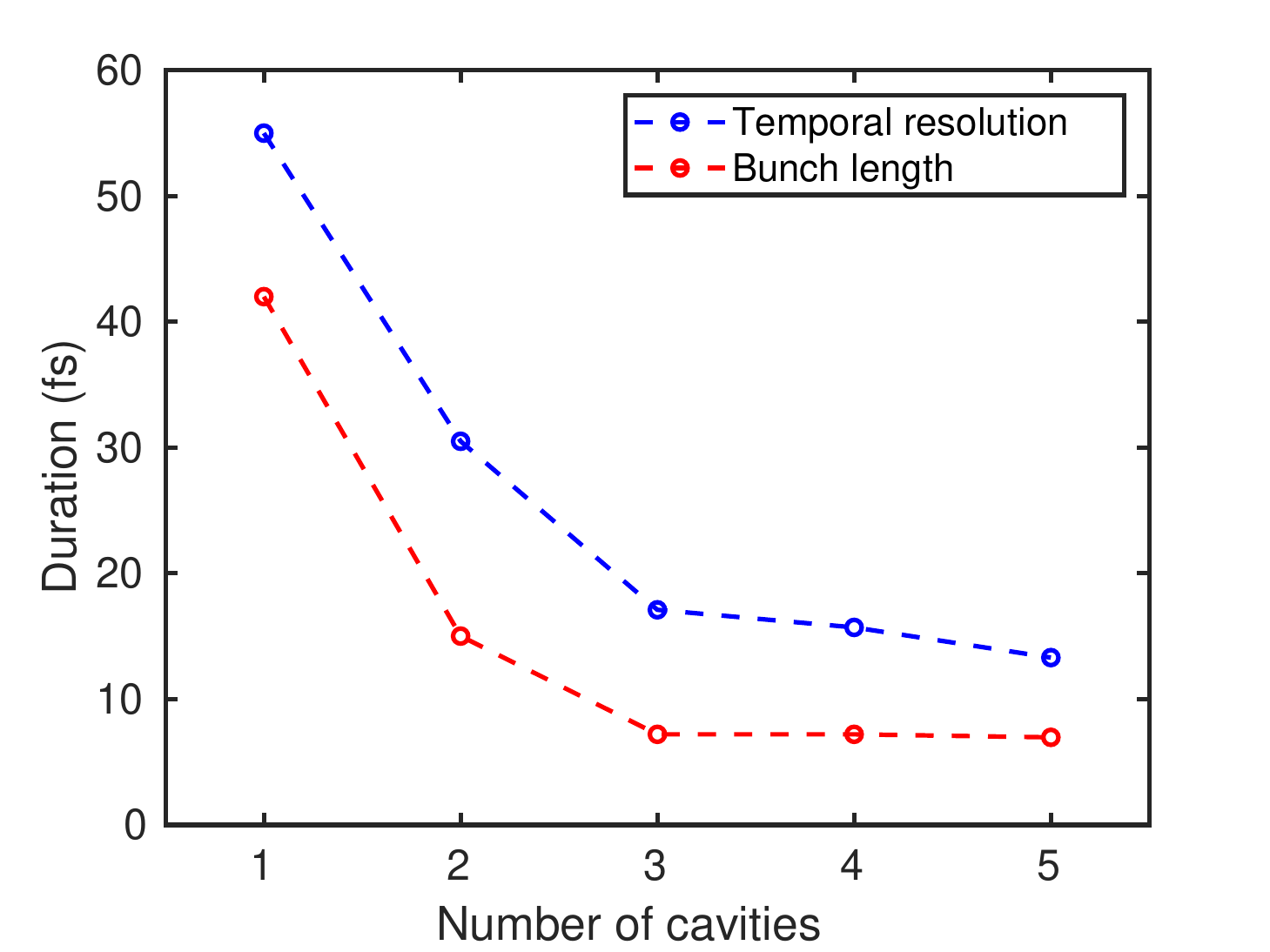}
        }
        \caption{(a) Trade-off between achievable temporal resolution and emittance for different numbers of SRF cavities. (b) Temporal resolution and bunch length for different numbers of SRF cavities at a constant emittance of 4 nm.}
        \label{FIG:stability_summary}
    \end{center}
\end{figure}

It is not clear from that single example case to what degree each of the SRF cavities is required in order reach that level of performance. To better understand that, we performed additional MOGA optimizations, this time optimizing the temporal resolution (including stability and bunch length). We did this five times, each time reducing the number of SRF cavities by one, while keeping the rest of the beamline unchanged. In Fig. \ref{FIG:stability_fronts}, the Pareto fronts of these optimizations are shown. In Fig. \ref{FIG:stability_lines} a cut along these fronts at an emittance of 4 nm is shown, and both the temporal resolution and the bunch length are plotted. From these plots, it is clear that in single shot mode at least three independent SRF cavities are needed for best performance, and it is primarily the bunch length that limits the temporal resolution below that number. But, given the ultimate single electron performance of this machine is well below one femtosecond, this mode of operation would require non-destructive high-precision time of arrival measurements for time-stamping to circumvent the effects of machine stability.

\begin{table}[!htbp] 
\caption{Sensitivity to Machine Parameters}
\begin{tabular}{l | c | c | c | c}
\hline
\hline
Parameter & Setting & Stability & Units & $\sigma_{t_0}$ (fs)\\
\hline
Gun amplitude    & 300  & 0.03  & kV & 2.3 \\
Buncher amplitude  & 26  & 0.001  & kV &  0.5 \\
Cavity 1 amplitude & 1150  & 0.03  & kV & 1.0 \\
Cavity 2 amplitude & 1170  & 0.03  & kV & 1.0\\
Cavity 3 amplitude & 1590  & 0.03  & kV & 0.5 \\
Cavity 4 amplitude & 1290  & 0.03  & kV & 0.4 \\
Cavity 5 amplitude & 1250  & 0.03  & kV & 0.2 \\
Buncher phase & -90  & .01  & deg. & 0.8 \\
Cavity 1 phase & -1  & .01  & deg. & 0.7 \\
Cavity 2 phase & -9  & .01  & deg. & 2.7 \\
Cavity 3 phase & -51  & .01  & deg. & 6.5 \\
Cavity 4 phase & -45  & .01  & deg. & 6.2 \\
Cavity 5 phase & -104  & .01  & deg. & 4.7 \\
\hline
Total &  &  &  & 13 \\
\hline
\hline
\end{tabular}
\label{tab:machine_settings}
\end{table}

\section{Conclusion}

We have presented a design for a flexible, high repetition rate MeV ultrafast electron diffraction apparatus based on the injector at Cornell University, and have characterized its performance from the space charge-free stroboscopic regime to the space charge-dominated regime pertinent to single shot diffraction. 

In the single electron per pulse regime, we find that the ultimate bunch length achievable in simulation is well below 1 fs rms, owing to the longitudinal phase space linearization that is possible using multiple accelerating and bunching cavities. In the case of very small spot sizes, the single electron/pulse bunch length is ultimately limited by transverse to longitudinal coupling. In practice, however, even state of the art RF synchronization produces time of arrival variations at the level of 10 fs rms or larger. We show that while the increased complexity of our RF system greatly reduces bunch length, it is unable to significantly reduce this arrival time instability. Considering the large gap between simulated bunch length performance and practical arrival time uncertainty, our work suggests that non-invasive time-stamping techniques with high temporal resolutions may be required to unlock attosecond MeV single electron pulses for experimental use.

With $10^5$ electrons per pulse, we find that the use of a small collimating aperture enables the use of strong transverse focusing to generate a dense transverse core with a large halo, wherein the halo is ultimately removed by a pinhole. Ultimately the system achieves 2-3 nm emittance (depending on photocathode MTE) and 10 fs rms pulse lengths; shorter pulses down to 5 fs rms are possible with 4-5 nm emittance. 

Finally, we analyze the ultimate temporal resolution of the device accounting for both bunch length and time of arrival variations using the measured field stability values in the Cornell SRF photoinjector. We do this as a function of the number of accelerating cavities included in the cryomodule to aid in future designs. We find that at least 3 cavities are required for significant performance gains. With 5 cavities, we show that field stability has a minor impact on high charge performance, yielding a total temporal resolution of $14$ fs rms.   

\begin{acknowledgments}
This work was supported by DOE Award No. DE-SC0021037 and by DOE Award No. DE-SC0020144.
\end{acknowledgments}

\section{References}

\bibliography{bibliography}





\end{document}